\documentclass[10pt,twocolumn]{article}
\usepackage{graphicx} 

\title{LLMs + Security = Trouble}

\usepackage[dvipsnames]{xcolor}
\usepackage[small]{titlesec} 
\usepackage{hyperref}
\usepackage{xspace}
\usepackage{url}

\hypersetup{
  colorlinks=true,
  linkcolor=MidnightBlue,
  citecolor=ForestGreen,
  urlcolor=RoyalBlue,
  filecolor=RoyalBlue,
  pdftitle={LLM + Security = Trouble},
  pdfauthor={Benjamin Livshits}
}

\newcommand{\point}[1]{\medskip\par\noindent\textbf{#1}.}
\author{Benjamin Livshits\\Imperial College London}
\date{\today}

\begin{document}
\sloppy

\maketitle

\begin{abstract}
  We argue that when it comes to producing secure code with AI, the prevailing ``fighting fire with fire'' approach---using probabilistic AI-based checkers or attackers to secure probabilistically generated code---fails to address the long tail of security bugs. As a result, systems may remain exposed to zero-day vulnerabilities that can be discovered by better-resourced or more persistent adversaries.

While neurosymbolic approaches that combine LLMs with formal methods are attractive in principle, we argue that they are difficult to reconcile with the ``vibe coding'' workflow common in LLM-assisted development: unless the end-to-end verification pipeline is fully automated, developers are repeatedly asked to validate specifications, resolve ambiguities, and adjudicate failures, making the human-in-the-loop a likely point of weakness, compromising secure-by-construction guarantees.

In this paper we argue that stronger security guarantees can be obtained by enforcing security constraints \emph{during} code generation (e.g., via constrained decoding), rather than relying solely on post-hoc detection and repair. This direction is particularly promising for diffusion-style code models, whose approach provides a natural elegant opportunity for modular, hierarchical security  enforcement, allowing us to combine  lower-latency generation techniques with generating secure-by-construction code.

\end{abstract}
\section{Introduction}
\label{sec:intro}
In recent months, a growing body of quantitative evaluations has highlighted the relative \emph{insecurity} of LLM-generated code.
Recent empirical studies~\cite{zhuWhenSoftwareSecurity2025,pearceAsleepKeyboardAssessing2022} consistently find that code produced by modern large language models exhibits elevated rates of security issues compared to human-written baselines~\cite{daiRethinkingEvaluationSecure2025}. Yan et al.~\cite{yanGuidingAIFix2025}, for instance, report non-trivial vulnerability rates across multiple models on established benchmarks and show that even tool-guided repair can be inconsistent and sometimes introduces new problems. Similarly, Chong et al.~\cite{chongArtificialIntelligenceGeneratedCode2024} find that LLM-generated C solutions often omit defensive checks (e.g., bounds checks and overflow protections), leading to higher vulnerability findings under static analysis; Kharma et al.~\cite{kharmaSecurityQualityLLMGenerated2025} further show that security posture varies by language and library choices, with models frequently defaulting to outdated or insecure practices. The reader is referred to Section~\ref{sec:insecure-code-generated} for more discussion.

Taken together, these results suggest that security vulnerabilities are not just another form of technical debt, but a systematic failure mode of AI-assisted code generation.
At the same time, LLM-based tools are increasingly capable of finding 0-day vulnerabilities, even in codebases that have already been heavily fuzzed\footnote{Evaluating and mitigating the growing risk of LLM-discovered 0-days: \url{https://red.anthropic.com/2026/zero-days/}} or scrutinized by expert human penetration testers\footnote{XBOW - The road to Top 1: How XBOW did it: \url{https://xbow.com/blog/top-1-how-xbow-did-it}}. While our evidence of the ability of LLMs to detect exploitable vulnerabilities in manually-written code is growing, there is every reason to expect the same to be true of generated code or of codebases that mix human- and machine-produced code.


\subsection{Paper Organization}
This paper is organized as follows.
Section~\ref{sec:fire} articulates why ``fighting fire with fire''---using probabilistic AI checkers/attackers to secure AI-generated code---is intrinsically limited by coverage constraints and the rarity of exploitable paths.
Section~\ref{sec:solutions} surveys solution directions that aim to provide stronger assurances than best-effort probabilistic reasoning, including neurosymbolic verification and structured/constraint-based generation techniques.
Finally, Section~\ref{sec:related} situates our discussion in the broader literature on insecure code generation, LLM-based vulnerability detection, and agentic security guardrails.

\section{Fighting Fire with Fire}
\label{sec:fire}

While the topics of prompt injection and directly malicious models are interesting, let us start with the pragmatic approach of finding and fixing security vulnerabilities.

\point{AI security triangle}
The AI \emph{security triangle} is the general patterns we've seen emerge, constisting of the following three \emph{roles}, as illustrated below.

\begin{itemize}
  \item \emph{generator}: this is the code producer, which may rely on a LLM, SLM, or some other code generation or synthesis approach;
  \item \emph{checker}: this is what analyzes the code, often using LLMs, but can also use more traditional tooling such as static analysis;
  \item \emph{attacker}: an approach that uses LLMs or a combination of these and manual attack generation approaches such as penetration testing, etc.
\end{itemize}
These roles sometimes become the basis for an \emph{agentic workflow}. Specific details vary from project to project; for instance, sometimes the generator and the checker are largely intervined~\cite{saulSCGAgentRecreatingBenefits2025}.
However, to summarize our fundamental argument: in real-world systems, the vulnerabilities that matter most are usually low-probability, high-impact events that frequently sit behind long and brittle chains of conditions. This is the reason why both real-world exploitation and competitive exploitation such as CTF events and HackerOne-style bug bounty sites place such a high financial premium on reproducible working exploits.

\point{State space explosion}
Large codebases have enormous state spaces and control-flow complexity, so any analysis technique---whether it is a traditional static analyzer, a fuzzing campaign, a test suite synthesized by an LLM, or an ``AI red team'' that proposes candidate exploits---runs into some version of path explosion: there are simply too many feasible executions, input shapes, environment configurations, and dependency interactions to explore exhaustively. This is especially acute for exploitability, because many practical exploits correspond to \emph{exceptional} execution paths that are intentionally rare in normal operation or through testing: error-handling branches, corner-case parsing logic, unusual privilege boundaries, or race windows that require precise timing and system load.

As a result, defenders who ``fight fire with fire'' by running LLM-based detection tools~\cite{chongArtificialIntelligenceGeneratedCode2024,heLargeLanguageModels2023} or deploying an automated attacker are typically optimizing for coverage under a fixed budget (time, compute, tool calls, rate limits), which means the attacker must necessarily miss a large fraction of the exceptional behaviors where high-value bugs hide. Worse, the defender's attacker is constrained by the need to be safe, reproducible, and non-disruptive (e.g., it cannot just freely DoS production  deployments or take destructive actions), while real adversaries can focus narrowly on one target, iterate indefinitely, and exploit any partial signal leaked by crashes, logs, or timing differences.

\point{Absence of issues is impossible to prove}
This asymmetry implies that even if an AI-based attacker succeeds in finding \emph{some} issues, its failure to find a given exploit provides pretty weak evidence of its absence: ``not found'' often just means ``not reached.'' Consequently, deploying an automated attacker does not eliminate the underlying risk; it merely changes which slice of the vulnerability space is explored, and a different attacker---human or AI-based---can still succeed by searching a different region or by investing more effort into the rare, exceptional paths that defensive testing is least likely to hit. The consequence of this is that an attacker with
\begin{itemize}\itemsep=-2pt
  \item a more powerful model;
  \item a more powerful hardware;
  \item or some insight that guides exploration of the code base
\end{itemize}
will often have the upper hand. Given LLM's perpensity to generate a lot of code and the high ROI of exploitation, this is especially true.

\point{Lack of forward security}
Troublingly, AI-generated code may have very weak \emph{forward security}: code that was produced a year ago may not withstand aggressive security hammering by today's top-line LLMs; similarly, code that is produced today may not withstand exploitation by the models in use a year from now. Furthermore, not everything is public: a defender using public models for identifying vulnerabilities may be no match to an attacker using \emph{privately developed} models, potentially deployed on more powerful custom hardware.

\subsection{The Valley of Probabilistic Reasoning}
Recent projects have used the idea of \emph{prefix tuning}~\cite{heLargeLanguageModels2023} to increase the ratio of secure code in programs generated by \texttt{CodeGen-2.7B} from~59\% to~92\%. Prefix tuning effectively relies on the transformer attention mechanism to exert a long-term influence on the computations of subsequent hidden states, including the prompt and the code to be generated.

This \emph{steers} the LM to generate programs that adhere to a security property, captured using a security fix diff, such as inserting sanitizing code to fix potential cross-site scripting issues. However, as pointed out in Fu~et al.~\cite{fuConstrainedDecodingSecure2024}, prefix tuning may lead to deterioration in terms of functional correctness of generated code.

Other recent proposals~\cite{wangFinetuningVulnerabilityspecificLarge2026} that advocate applying static analysis to LLM-generated code inherint the traditional drawbacks of static analysis, including false positives and scalability challenges. Fuzzing-based testing enjoys an increased coverage as a result of LLM-based exploration, but this still falls short of full coverage for large code bases~\cite{zhangLowCostComprehensiveNontextual2025}.

\section{Beyond Probabilistic Guarantees}
\label{sec:solutions}

There are two main approaches people have proposed in the last several years that avoid the pitfalls of the AI security traiangle we described above.

\subsection{Neurosymbolic Reasoning}
Neuro-symbolic code generation is a hybrid approach that integrates the statistical pattern-matching capabilities of neural networks such as LLMs with the rigorous, logical structure of symbolic computation and programming language (PL) theory. This integration aims to bridge the \emph{semantic gap} inherent in standard LLMs, elevating code generation from probabilistic text prediction to reliable, verifiable, and structurally sound software synthesis. While this direction pursues a number of laudable goals, we feel that there's a fundamental disconnect between the mentality of ``vibe coding'' common is LLM-based code production and the rigour that is needed for the developer to be part of the code reasoning pipeline, unless the end-to-end neurosymbolic approach is fully automated. If not, the human-in-the-loop becomes the weakest link, as we highlight below, using two recent projects as examples.

\begin{itemize}
  \item \textbf{Astrogator.}
    Councilman et al.~\cite{councilmanFormalVerificationLLMGenerated2025b} introduce \textsc{Astrogator}, a system designed to provide formal correctness guarantees for code generated by LLMs. While LLMs excel at generating code from natural language, they often produce subtle bugs or hallucinations that make them unsuitable for mission-critical systems. \textsc{Astrogator} addresses this by introducing a Formal Query Language~(FQL). The FQL acts as a bridge: it is precise enough for formal verification but sufficiently high-level and natural for a user to review and confirm it matches their intent.
    The system translates both the user's formal query and the LLM-generated code into a novel State Calculus---a simplified mathematical model that tracks system changes like file creations or package installations. A symbolic interpreter then compares the behaviors of the query and the code to ensure they match.
    In an evaluation of~1,260 programs across~21 tasks, \textsc{Astrogator} successfully verified correct code in~83\% of cases and identified incorrect code in~92\% of cases. The authors highlight that a key innovation is the use of a Knowledge Base to map high-level concepts (e.g., \emph{Apache server}) to platform-specific details (\texttt{httpd} of Red Hat Enterpirse Linux), reducing the specialized knowledge required by the user.

  \item\textbf{ARC.}
    Bayless~\cite{baylessNeurosymbolicApproachNatural2025} present ARC, a two-stage neurosymbolic framework that~(1) uses LLMs with optional human guidance to formalize natural language policies, allowing finegrained control of the formalization process, and~(2) uses inference-time autoformalization to validate logical correctness of natural language statements against those policies. When correctness is paramount, the framework performs multiple redundant formalization steps at inference time, cross checking the formalizations for semantic equivalence, and it produces auditable logical artifacts that substantiate verification outcomes and can be used to improve the original text. Their benchmarks demonstrate that their system called ARC exceeds~99\% soundness, indicating a near-zero false positive rate in identifying logical validity; this high soundness is also reflected in conventional metrics such as false positive rate and precision, where ARC outperforms competing approaches. In addition, ARC delivers \emph{explainable} verdicts and provides actionable feedback that LLMs can utilize to refine their outputs, establishing a robust foundation for future research aimed at pushing assurance boundaries to three nines and beyond.
\end{itemize}
Both \textsc{Astrogator} and ARC show significant strides towards enhancing the quality of generated code. However, the developer still needs to get involved in a \emph{significant percentrage} of cases that fail automatic verification that cannot be definitely deemed as broken or buggy. Given the developer's propensity to ``accept all changes'' by default in the paradigm of vibe coding, this is not a match made in heaven.

\subsection{Security by Constuction, But How?}

Recent papers argue that non-autoregressive decoding is better than autoregressive decoding for the problem of controlled text generation under constraints~\cite{kumar2022gradientbasedconstrainedsamplinglanguage}. The same arguments hold for code generation under constraints. Indeed, during autoregressive decoding, we cannot evaluate the properties of the \emph{entire} program during the generation because only a \emph{partial program} is available.

For example, if the partially generated code has not sanitized untrusted user input \emph{yet}, it does not mean that the entire generated code would not sanitize untrusted user input at a later point in execution, so we cannot yet know whether the partial program is insecure. On the contrary, non-autoregressive decoding generates the entire program altogether, which enables us to evaluate constraints as well as enforce constraints over the whole program.

Diffusion models~\cite{gongDiffuCoderUnderstandingImproving2025, kapurDiffusionSyntaxTrees2024, liAutoregressionEmpiricalStudy2025, liDiffuGuardHowIntrinsic2025, zhangExploringPowerDiffusion2025,zengTreeDiffASTGuidedCode2025} have been heralded as an attractive alternative for code generation, largely due to their speed and ability to better exploit parallelism of more hierarchical code generation, compared to auto-regressive models. The reader is referred to Section~\ref{sec:diffusion-models} for more.

\subsection{Small World Hyphothesis for Security Fixes}

In diffusion model-based code generation, we have the opportunity to detect potential issues at the structural level of AST nodes such as modules, functions, loops, conditionals, and the like. While security errors may be scattered across a multitude of modules and source files, the key observation is that most security \emph{fixes} can be made \emph{locally}, in a manner that is contained within a single function call.

PatchDB~\cite{wangPatchDBLargeScaleSecurity2021} has a plethora of examples that confirm this claim, where most vulnerability fixes are only~1 or~2 lines long and are localized. This local fix nature is generally true of memory vulnerabilities, such as buffer overruns or double frees. It's also true of cross-site scripting or path traversale attacks.

We therefore see a lot of potential in building models that do \emph{constrained decoding}~\cite{fuConstrainedDecodingSecure2024, parkFlexibleEfficientGrammarConstrained2025, storhaugEfficientAvoidanceVulnerabilities2023} in the context of diffusion-based code generation. However, we emphasize that in order to create security guarantees, we need to have \emph{hard, non-probabilistic constraints} that limit what code is allowed to be produced. These can take the form of pro-active invariant checks such as inserted bounds check for array accesses (buffer overrun prevention) and sanitization for strings that are output to HTML (cross-site scripting prevention).

\newcommand{\sskip}{}

\section{Related Work}
\label{sec:related}

This section synthesizes recent evidence that LLM-assisted coding can introduce substantial security risk, and it surveys emerging techniques that attempt to reduce that risk through more principled evaluation and stronger control over generation. Across empirical studies and new benchmarks, the literature reports frequent insecure defaults, inconsistent self-repair, and serious validity concerns (e.g., moving targets and data leakage), motivating metrics that jointly measure functionality and security and the use of multiple analyzers to avoid overestimation.

In response, researchers have proposed mitigation pipelines for secure generation and repair that combine guidelines, retrieval, testing, and tool feedback, as well as correctness-focused mechanisms, such as constrained (or semantically guided) decoding and formal or symbolic verification. Complementary work uses LLMs as security instruments~---~for vulnerability detection, patch generation, and fuzzing/auditing workflows~---~and separately studies guardrails for agentic systems that can take tool-using actions under adversarial inputs.

Finally, we highlight broader modeling and reliability trends (e.g., structure-aware code representations, uncertainty estimation grounded in semantic equivalence, and diffusion-style code models) that push code LMs beyond surface-level pattern matching toward semantics-aware reasoning.

\subsection{Security Risks of LM Code Generation}
\label{sec:insecure-code-generated}
Zhu et al.~\cite{zhuSpecificationGuidedVulnerabilityDetection2025} provide a systematic survey of the intersection between software security and Large Language Models, detailing how these models have revolutionized traditional software analysis and testing. The authors deconstruct the application of LLMs into critical domains, including fuzzing, unit testing, program repair, bug reproduction, data-driven bug detection, and bug triage. A key contribution of this work is the deconstruction of these security techniques into a three-stage pipeline: pre-processing (data collection and fine-tuning), prompt generation (context extraction and instruction crafting), and post-processing (validation of generated code or inputs). This structural analysis allows researchers to pinpoint exactly where the generative capabilities of LLMs add value compared to traditional deep learning models.

The survey highlights that while traditional deep learning required massive amounts of high-quality, labeled data, LLMs can often perform security tasks effectively in zero-shot or few-shot settings through sophisticated prompt engineering. In fuzzing, for instance, LLMs are used to generate highly structured inputs for complex grammars or to synthesize fuzz drivers, while in bug reproduction, they utilize natural language capabilities to interpret bug reports. However, the authors also identify significant limitations, such as the high computational cost of large-scale deployment, the potential for hallucinations, and the lack of understanding of complex logic in \emph{unseen} code. They suggest that future research should focus on multi-step prompt generation and better post-processing validation to mitigate these risks. By reviewing the state-of-the-art across over 100 recent works, the study establishes a comprehensive roadmap for using LLMs to reduce the manual effort involved in maintaining software quality and security across the development lifecycle.

\sskip
Yan et al.~\cite{yanGuidingAIFix2025} conduct an extensive empirical study on the tendencies of both proprietary and open-weight Large Language Models to produce insecure code and their capacity to repair such vulnerabilities when guided by feedback. The study evaluates a broad range of models using established benchmarks like SecurityEval and SecCodePLT, focusing on Python. The researchers identify two primary paradigms for improving code security: proactive vulnerability prevention and post-hoc vulnerability repair. Proactive prevention involves guiding the model to avoid security pitfalls during the initial generation phase, often through the use of self-generated vulnerability hints. Post-hoc repair, meanwhile, occurs after a vulnerability is detected by an external tool like CodeQL, using the tool's output as feedback for a second round of generation.

The findings reveal that current LLMs generate vulnerable code at rates between~9.8\% and~42.1\%, with even the most advanced models frequently overlooking critical security practices. Interestingly, the study shows that while self-generated hints can reduce the incidence of vulnerabilities, they often contain irrelevant or incorrect information, limiting their reliability. In the post-hoc repair phase, the researchers compared ``direct feedback'' (raw CodeQL error messages) with ``explained feedback'' (GPT-4o summaries of the errors). Their results demonstrate that models with strong instruction-following capabilities significantly benefit from explained, contextualized feedback, which provides detailed suggestions and actions for resolution. The study concludes with actionable suggestions for developers, such as utilizing multi-round reasoning and external validators, to mitigate the risks of AI-assisted coding. This work underscores the importance of the human-in-the-loop workflow and the need for more nuanced guidance than simple binary error reports.

\sskip
Chong et al.~\cite{chongArtificialIntelligenceGeneratedCode2024} investigate the security and quality of code generated by large language models, specifically focusing on OpenAI's GPT-4o. The authors introduce the EXACT framework to evaluate over~200 LeetCode problems, algorithm implementations, and cryptographic functions written in C, contrasting these with human-written equivalents. A central finding is that while LLMs can solve many programming tasks, they often fail at complex ones and produce subtle functional errors, such as an incorrect SHA-1 implementation that compiles but yields wrong values. Security-wise, LLM-generated code consistently exhibits higher vulnerability rates, frequently due to a lack of defensive programming constructs like boundary checks for buffers or protections against integer overflows. Static analysis revealed that LLM-generated code contained~11.2\% more security issues in LeetCode tasks and~7.1\% more in algorithm tasks than human code. Furthermore, the paper evaluates \emph{feedback loops} where the model is prompted to fix its own bugs; however, this process is inconsistent, as the LLM frequently introduces new vulnerabilities into previously safe code or increases the cyclomatic complexity of the solution. The authors highlight a \emph{false sense of productivity} provided by AI tools, where developers may focus on task completion time rather than the long-term maintainability and security of the \emph{bare-bones} code produced. They advocate for a more cautious adoption of AI in security-critical software engineering, noting that prompt engineering alone does not guarantee secure outputs and can potentially exacerbate existing flaws.

\sskip
Sallou et al.~\cite{sallouBreakingSilenceThreats2024} address the growing body of research on Large Language Models in software engineering, warning of significant \emph{threats to validity} that often go unaddressed. The authors categorize these threats into three main areas: the use of closed-source models, implicit data leakage, and reproducibility challenges. Closed-source models like ChatGPT represent a \emph{moving target}; because they are updated frequently without public documentation, results obtained at one point in time may be impossible to replicate later. This \emph{evolution unpredictability} makes it difficult for researchers to distinguish between improvements caused by their own methodologies and those resulting from underlying model updates. Data leakage is another critical concern, as LLMs are trained on massive public datasets that likely include the benchmarks used to evaluate them. This can lead to a \emph{blurry separation} between training and testing data, where the model appears to solve problems through memorization rather than true reasoning. To combat these issues, the paper proposes guidelines such as using \emph{metamorphic testing} or code obfuscation to create new test cases, and providing extensive metadata to improve traceability. The authors suggest that researchers should favor open-source models whenever possible to ensure long-term reproducibility. By \emph{breaking the silence} on these methodological pitfalls, the paper aims to foster a more rigorous and scientifically sound approach to evaluating AI in software engineering.

\sskip
Zhao et al.~\cite{liangGrammarBasedCodeRepresentation2025} investigate the security of \emph{vibe coding}, a paradigm where developers use natural language to direct AI agents to complete complex tasks with minimal supervision. To evaluate this, they introduce the SUSVIBES benchmark, which consists of~200 software engineering tasks derived from real-world open-source vulnerabilities. Unlike previous benchmarks that focused on single functions, SUSVIBES requires agents to edit multiple files across large codebases, reflecting the actual complexity of modern software maintenance. The results of their study are \emph{disturbing}: while high-performing agents like SWE-Agent (using Claude 4 Sonnet) could solve~61\% of tasks functionally, only~10.5\% of those solutions were secure. This gap indicates that agents frequently produce code that passes unit tests but leaves the system open to historical vulnerabilities. The authors demonstrate that simple mitigation strategies, such as adding a generic security reminder to the prompt, are largely ineffective and can even degrade functional performance. The study also identifies that different models have different \emph{blind spots} regarding Common Weakness Enumerations (CWEs). The paper highlights a qualitative example where an agent implements a password verification function correctly but fails to include timing-attack protections, creating a side-channel vulnerability. By exposing these risks, Zhao et al. argue for more sophisticated safety strategies that proactively identify and mitigate vulnerabilities at the repository level.

\sskip
Chen et al.~\cite{chenSecureAgentBenchBenchmarkingSecure2025} introduce SECUREAGENTBENCH, a benchmark designed to evaluate the secure code generation capabilities of LLM-powered agents in realistic scenarios. They contend that previous benchmarks are insufficient because they simplify tasks to function-level completion and ignore the original context of vulnerability introduction. SECUREAGENTBENCH provides~105 coding tasks requiring multi-file edits in large repositories, grounded in real-world vulnerabilities from the OSS-Fuzz project. The benchmark assesses functional correctness through differential testing, checks for historical vulnerabilities using proof-of-concept (PoC) exploits, and uses static application security testing (SAST) tools to detect newly introduced security risks. Evaluating three representative agents—SWE-agent, OpenHands, and Aider—the authors find that all struggle significantly; even the best-performing combination resolved only~15.2\% of tasks correctly and securely. Notably, the study finds that agents often introduce new vulnerabilities not present in the original code, with over~20\% of functionally correct solutions containing such flaws. Like other studies, the authors observe that explicit security reminders in prompts offer negligible improvements. The benchmark serves as a critical resource for developing more reliable AI-driven development tools that can handle project-wide dependencies and security principles.

\sskip
Andersson et al.~\cite{anderssonPoCoAgenticProofofConcept2025} introduce PoCo, an agentic framework designed to automate the generation of executable Proof-of-Concept (PoC) exploits for smart contract vulnerabilities. In the high-stakes environment of Web3, manually creating PoCs—which demonstrate that a reported vulnerability is genuine—is a time-consuming and error-prone process for auditors. PoCo addresses this by taking a natural-language description of a vulnerability and autonomously crafting an executable exploit compatible with the Foundry testing framework. The framework utilizes a \emph{Reason-Act-Observe} loop (ReAct), where an LLM agent interacts with domain-specific tools for compilation and testing. Unlike simple prompting methods, PoCo can iteratively refine its exploit code based on execution feedback, fixing compilation errors or logical flaws in its attack strategy. Evaluating the system on the Proof-of-patch dataset, the authors show that PoCo consistently outperforms zero-shot baselines, successfully generating correct PoCs for a majority of cases. The authors emphasize that PoCo’s primary contribution is providing developers with \emph{actionable test cases} to reproduce and fix issues rapidly. By grounding the agent’s behavior in established smart contract auditing best practices, PoCo demonstrates the potential of agentic AI to enhance the efficiency of security reviews in the blockchain ecosystem.

\subsection{Vulnerability Detection with LLMs}

Saul et al.~\cite{saulSCGAgentRecreatingBenefits2025} introduce SCGAgent, an agentic framework designed to improve the security of LLM-generated code without sacrificing functional correctness. They argue that traditional methods, such as simple security reminders or fine-tuning, are often insufficient because they either provide marginal improvements or require access to proprietary model weights. SCGAgent mimics a \emph{junior developer} training approach by using detailed, expert-written secure coding guidelines. The framework operates through a structured workflow: it first predicts the potential Common Weakness Enumerations (CWEs) relevant to a task, retrieves specific guidelines to address them, and then iteratively modifies the generated code to follow these instructions.

To ensure that security constraints do not break functionality, SCGAgent incorporates an \emph{Enforce-Functionality} module. This module generates automated unit tests and uses them to validate each revision of the code. If a test fails, the agent determines whether the flaw lies in the code or the test itself and repairs it accordingly. Evaluation on the CWEval benchmark demonstrates that SCGAgent significantly boosts security—increasing the percentage of secure and functional samples from~61\% to~76\% when using Claude Sonnet-3.7 as a base. Interestingly, the authors find that SCGAgent allows non-reasoning models to match or exceed the performance of more computationally expensive reasoning models. This modular approach allows for easy expansion as new vulnerabilities are discovered, providing a scalable path toward trustworthy AI-assisted programming.

\sskip
Nong and Cheng~\cite{nongAPPATCHAutomatedAdaptive2025} introduce APPATCH, a framework designed to improve the performance of Large Language Models in the context of real-world software vulnerability patching. The authors argue that while LLMs show promise, they often fail to generate correct patches due to a lack of guidance in reasoning about complex code semantics. To address this, APPATCH utilizes a technique called semantics-aware scoping, which uses interprocedural backward slicing to isolate the specific code statements relevant to a vulnerability's root cause. By narrowing the model's focus to these essential fragments, the system avoids the token limits and distractions associated with large codebases.

A core innovation of the paper is the dynamic adaptive prompting strategy, which automatically selects relevant exemplars from a pre-mined database of successful patches. This database is constructed offline by leveraging LLMs to generate reasoning steps for known vulnerabilities, which then serve as Chain-of-Thought (CoT) guides for the model during the actual patching process. The system also includes a multi-faceted validation phase where an ensemble of LLMs cross-verifies candidate patches for functional correctness and security. Evaluation on zero-day and real-world vulnerability datasets demonstrated significant improvements over baseline zero-shot prompting and traditional non-LLM techniques, achieving up to a~28\% increase in F1-score. The authors conclude that guiding LLMs through structured semantic reasoning is critical for moving beyond simple code completion toward reliable security remediation.

\sskip
In Luo et al.~\cite{luoGuidingLLMbasedSmart} the researchers proposed FSM-SCG, a novel framework designed to enhance the effectiveness and security of smart contract generation using LLMs. Recognizing that direct code generation often leads to syntax errors and security vulnerabilities due to the low-resource nature of blockchain languages like Solidity, the authors introduced an intermediate representation called SmartFSM. This enhanced finite state machine abstracts user requirements into states, variables, and transitions, which then guide the LLM through a ``requirement-to-FSM-to-code'' (R2F2C) generation pattern. To further ensure reliability, the framework incorporates an automated feedback loop that utilizes compilation results and security checks from tools like Slither to iteratively refine the generated code.

Experimental results demonstrate that FSM-SCG significantly outperforms existing methods across various open-source and closed-source LLMs. By fine-tuning models like Llama-3.1-8B on a custom-built dataset of~30,000 items, the framework achieved a compilation success rate of~95.1\%, marking a~48\% improvement over the best baseline. Additionally, the approach substantially improved security, reducing the average vulnerability risk score by~68\%. Ablation studies confirmed that both the structured SmartFSM guidance and the iterative feedback mechanism are critical components for generating complex, functional, and secure smart contracts.

\sskip
Hajipour et al.~\cite{hajipourHexaCoderSecureCode2024} present HexaCoder, a novel methodology aimed at enhancing the security of code generated by LLMs through oracle-guided synthetic data and a specialized two-step generation process. The research addresses the critical shortage of high-quality training data that pairs vulnerable code with its secure counterparts. HexaCoder solves this by using a security oracle (specifically CodeQL) to identify vulnerabilities in code generated by various models, which are then repaired by a high-capacity LLM using the oracle's feedback. This results in a synthetic dataset of vulnerable-fixed pairs that includes the necessary security-related libraries often missing in standard datasets. The models are fine-tuned using Low-Rank Adaptation (LoRA) on this data. Furthermore, the paper introduces a two-step inference method: the model first predicts the required security libraries and headers based on the prompt, and then completes the main logic. This approach prevents the model from being ``locked in'' to insecure default libraries. Experimental results across multiple benchmarks showed that HexaCoder could reduce the number of vulnerable code instances by up to~85\% while maintaining functional correctness. The authors highlight that integrating external security knowledge via oracles during training allows smaller models to achieve security levels comparable to much larger, proprietary counterparts.

\sskip
Bappy et al.~\cite{bappyCaseStudyFinetuning2025a} explore the feasibility of using Small Language Models (SLMs) for accurate and private Common Weakness Enumeration (CWE) detection in Python. They focus on organizations with strict data governance policies, such as those in finance or healthcare, who cannot send proprietary code to cloud-based LLM providers. The study investigates whether a~350-million parameter model, codegen-mono, can be effectively fine-tuned to detect the MITRE Top 25 CWEs. To overcome the scarcity of labeled Python security data, the authors developed a semi-supervised dataset generation pipeline using a reasoning-focused model (Gemini-2.0-flash-thinking) followed by rigorous human review. This process resulted in a high-quality dataset of 500 examples covering diverse vulnerability types. While the base SLM initially failed to identify any CWEs, instruction-following fine-tuning transformed its performance. The specialized SLM achieved near-perfect metrics on the test set, including 99\% accuracy and 100\% recall. The findings suggest that when properly fine-tuned on targeted security data, SLMs can provide a resource-efficient and privacy-preserving alternative to cloud-hosted models. The authors emphasize that this approach allows advanced security analysis to be integrated directly into local development workflows without compromising intellectual property or incurring high API costs.

\sskip
Wang et al. (2026)~\cite{wangFinetuningVulnerabilityspecificLarge2026} propose HyVD-VP, a hybrid vulnerability detection method specifically tailored for large-scale enterprise Java projects. The authors note that existing LLM-based approaches often struggle with the long-range dependencies and complex control flows typical of enterprise software. To solve this, they first fine-tune a specialized model called VulDetLLM using a two-stage process that includes Supervised Fine-Tuning (SFT) and Direct Preference Optimization (DPO) to align the model with expert security judgment. The HyVD-VP system addresses the context-window bottleneck by using Control Flow Graph (CFG) based semantic segmentation to break down long programs into meaningful slices. These slices are enhanced with Retrieval-Augmented Generation (RAG), which pulls relevant vulnerability signatures from a local database. A major distinction of this work is the integration of a Program-Assisted Language (PAL) model that generates lightweight test harnesses to collect dynamic runtime evidence. Finally, an LLM-as-a-Judge module reconciles the static detection results from VulDetLLM with the dynamic evidence from the PAL module to make a final, explainable decision. The method achieved a 96.3\% accuracy on Java datasets and successfully identified 16 previously unknown vulnerabilities in real-world enterprise projects, demonstrating its industrial readiness and ability to reduce false positives through multi-modal validation.

\sskip
Kaniewski et al.~\cite{kaniewskiSystematicLiteratureReview2025} provide a comprehensive Systematic Literature Review (SLR) analyzing the rapidly evolving landscape of LLM-based software vulnerability detection from 2020 to mid-2025. By reviewing 227 studies, the authors establish a detailed taxonomy that categorizes research across four dimensions: task formulation (binary vs. multi-class classification), input representation (raw text vs. structure-aware graphs), system architecture (LLM-centric vs. hybrid), and adaptation techniques (prompting vs. fine-tuning). The review highlights a significant shift from simple binary classification toward more complex reasoning about vulnerability root causes and localization. It also notes an increasing trend in using structure-aware inputs, such as Code Property Graphs (CPGs) or program slices, to overcome the limitations of plain-text representations. A critical contribution of this SLR is its in-depth analysis of datasets, revealing persistent issues with class imbalance, label noise, and limited CWE coverage in existing benchmarks. The authors identify a ``fragmented research landscape'' where inconsistent evaluation protocols make it difficult to compare performance across studies. They advocate for the adoption of more realistic, project-level benchmarks and standardized metrics to improve the reproducibility and transparency of the field. This review serves as a roadmap for future research, pointing toward agentic workflows and multi-task learning as key areas for growth.

\sskip
Safdar et al.~\cite{safdarDataContextMatter2025} address the fundamental problem of generalizability in AI-based software vulnerability detection. They observe that many existing models perform exceptionally well on benchmark datasets but fail when applied to unseen, real-world codebases. To investigate this, they introduced VulGate, a high-quality, rigorously curated dataset containing over 236,000 function-level samples. VulGate unifies several established corpora and integrates ``hard negative'' samples—function pairs with high semantic similarity where one is vulnerable and the other is not. These hard negatives are designed to force models to learn subtle semantic patterns rather than relying on superficial syntactic cues. The authors benchmarked various encoder-only and decoder-only architectures, finding that encoder-based models like UniXcoder, when combined with extended context windows (1024 tokens), significantly outperform other models in generalization. Their top-performing configuration achieved a 6.8\% improvement in recall on the BigVul benchmark and showed a remarkably low performance drop (only 4-6\%) when transitioning to completely unseen datasets. The study concludes that both the quality of training data—specifically the inclusion of diverse, cleaned samples—and the choice of model architecture are critical factors in building robust systems that can operate effectively across different software ecosystems.

\sskip
Fei et al.~\cite{feiLargeLanguageModels2025} conduct the first systematic evaluation of LLMs for JavaScript vulnerability detection, introducing the ARENAJS benchmark and the FORGEJS generation framework. The authors highlight that JavaScript presents unique challenges due to its prevalence in both frontend (browser) and backend (Node.js) environments, each with distinct vulnerability patterns like prototype pollution or DOM-based XSS. They identify three major defects in existing benchmarks: incomplete CWE coverage, underestimation due to rigid string matching, and overestimation caused by using isolated code snippets rather than complete projects. FORGEJS addresses these by aggregating data from heterogeneous sources, covering 218 CWE types, and constructing realistic, project-level evaluation pairs. The study's results are sobering: they reveal that while models often ``correctly'' identify vulnerabilities in snippets, their performance drops significantly in project-level contexts where they must navigate complex dependencies and taint flows. The authors found that leading models frequently rely on ``surface features'' like filenames or comments rather than true semantic understanding. They introduce the VD-S metric to evaluate deployment readiness under engineering constraints and conclude that current LLMs are not yet reliable enough for production-level JavaScript auditing, as they miss over~75\% of real vulnerabilities when false positive rates are strictly controlled.

\sskip
Hajipour et al.~\cite{hajipourHexaCoderSecureCode2024} present \textsc{HexaCoder}, a novel methodology aimed at enhancing the security of code generated by LLMs through oracle-guided synthetic data and a specialized two-step generation process. The research addresses the critical shortage of high-quality training data that pairs vulnerable code with its secure counterparts. \textsc{HexaCoder} solves this by using a security oracle (specifically CodeQL) to identify vulnerabilities in code generated by various models, which are then repaired by a high-capacity LLM using the oracle's feedback. This results in a synthetic dataset of vulnerable-fixed pairs that includes the necessary security-related libraries often missing in standard datasets. The models are fine-tuned using Low-Rank Adaptation (LoRA) on this data.

Furthermore, the paper introduces a two-step inference method: the model first predicts the required security libraries and headers based on the prompt, and then completes the main logic. This approach prevents the model from being ``locked in'' to insecure default libraries. Experimental results across multiple benchmarks showed that \textsc{HexaCoder} could reduce the number of vulnerable code instances by up to~85\% while maintaining functional correctness. The authors highlight that integrating external security knowledge via oracles during training allows smaller models to achieve security levels comparable to much larger, proprietary counterparts.

\subsection{Improving the Security of LM Code}
ARC by Bayless~et al.~\cite{baylessNeurosymbolicApproachNatural2025} distinguishes itself from traditional LLM evaluation methods by moving away from probabilistic judging and toward mathematical verification. The framework's core strength lies in its ability to generate auditable logical artifacts that provide mathematical proofs for why an answer is accepted or rejected. By formalizing policies into SMT-LIB, the system creates a definitive source of truth that remains consistent regardless of the LLM's internal fluctuations. This allows organizations to resolve policy ambiguities offline with human experts, ensuring the guardrail's logic is both precise and legally defensible before it ever interacts with a customer.

One significant pitfall identified in the research is the intentional sacrifice of recall for the sake of soundness. ARC is designed to be highly conservative; if the system is not entirely certain about a translation, it will flag it as \texttt{TranslationAmbiguous} rather than risk a false approval.

In benchmarks, ARC's high soundness of over~99\% resulted in a recall of only~15.6\%. While other methods like RefChecker achieved~87.6\% recall, they did so at the cost of soundness dropping to~84.4\%, making them less suitable for safety-critical domains where false approvals are more costly than false rejections. For businesses, this conservative bias means many valid answers might be initially rejected or held for human review.

The authors also highlight several technical and scaling pitfalls that affect current deployment. Complex policies still require a significant upfront investment in human vetting to resolve contradictions or implicit assumptions that automated tools may miss. The use of redundant translation, which involves calling three LLMs simultaneously to check for consensus, adds~5 to~15 seconds of latency and increases API costs per query. Furthermore, documents that rely heavily on numerical tables, cross-references, or complex nested negations can lead to \texttt{TooComplex} findings where the logic exceeds the system's current processing limits. As policies grow to hundreds of pages, the resulting models can contain thousands of rules, making manual vetting and maintenance increasingly difficult for human experts.

\sskip
Miculicich et al.~\cite{miculicichVeriGuardEnhancingLLM2025} present \textsc{VeriGuard}, a novel framework that aims to provide formal safety guarantees for LLM-based agents. The authors argue that existing safety mechanisms, such as input/output filtering, are insufficient because they rely on pattern matching and cannot cover the dynamic state space of autonomous agents. VeriGuard addresses this by shifting to a \emph{correct-by-construction} paradigm. The framework operates in two main stages: offline policy generation and online action monitoring. During the offline stage, \textsc{VeriGuard} takes a natural language security request and an agent specification to synthesize a behavioral policy function paired with \emph{formal reasoning constraints}, such as pre- and post-conditions.

These artifacts are then subjected to a rigorous iterative refinement loop that involves automated unit testing and formal verification using the Nagini verifier. If verification fails, the verifier provides a specific counterexample or logical inconsistency, which the agent uses to refine the policy until it is provably correct. In the second stage, the verified policy serves as a runtime monitor that intercepts every action proposed by the agent, ensuring it complies with the verified safety contract before execution. The paper demonstrates that \textsc{VeriGuard} can prevent harmful actions—such as unauthorized data exfiltration—that traditional filters might miss. By integrating formal methods into the agent's workflow, \textsc{VeriGuard} provides a robust safeguard for autonomous systems in mission-critical environments.

Licorish et al.~\cite{licorishComparingHumanLLM2025} provide a comprehensive empirical comparison between human-generated code and the outputs of Large Language Models (LLMs), specifically focusing on GPT-4. The study utilizes a benchmark dataset of~72 distinct Python software engineering tasks to evaluate four critical quality dimensions: adherence to coding standards, security vulnerabilities, code complexity, and functional correctness. To maintain objective standards, the researchers employed industry-standard static analysis tools including \emph{Pylint} for quality, \emph{Radon} for complexity, and \emph{Bandit} for security analysis, complemented by \emph{Pytest} for verifying semantic success.

The results of the study reveal a complex trade-off between human expertise and automated generation. While GPT-4 achieved a significantly higher functional pass rate of~87.3\% compared to~54.9\% for a fourth-year computing student, the human-generated code consistently outperformed the LLM in adhering to Python coding standards. Humans achieved an average \emph{Pylint} score of~9.8, whereas GPT-4 averaged~9.1, often failing in areas such as documentation and docstring completeness. In terms of security, both sources exhibited vulnerabilities like hard-coded API keys and insecure subprocess handling.

However, the paper shows that~60\% of the security issues in LLM-generated code were of high severity, compared to~45\% in human code. Furthermore, LLM-generated code was found to be approximately~61\% more complex on average (with a mean cyclomatic complexity of~5.0 versus~3.1 for humans), suggesting a tendency toward over-engineering. The authors conclude that while LLMs are proficient for well-defined tasks, human programmers remain superior for complex problem-solving that requires deep domain knowledge, innovative solutions, or unconventional thinking.

\sskip
Fu et al.~\cite{fuConstrainedDecodingSecure2024} investigate the critical intersection of security and functional correctness in automated code generation. They argue that previous security-focused defenses, such as prefix tuning, often give a ``false sense of security'' by generating code that is secure but functionally useless. To address this, they introduce CodeGuard+, a benchmark consisting of~91 prompts covering~34 Common Weakness Enumerations~(CWEs) in C, C++, and Python, alongside a new metric called \emph{secure-pass@k}. This metric evaluates the likelihood that a model produces code that is both semantically correct and free of vulnerabilities.

The researchers study the impact of decoding strategies and propose a ``secure-by-construction'' approach through \emph{Constrained Beam Sampling}. This technique enforces positive security constraints—such as requiring safe library functions (e.g., \emph{snprintf} instead of \emph{sprintf}) or mandatory array index bound checks—directly during the inference process without requiring specialized training data. Their evaluation of eight state-of-the-art models demonstrates that constrained decoding is more effective than the previous state-of-the-art prefix tuning  called \textsc{Sven}. Notably, they found that open-source models using constrained decoding could outperform the unconstrained output of GPT-4 in generating secure and correct code. The study highlights that code LLMs are highly sensitive to the choice of decoding technique, and using sampling-based constrained methods allows for greater exploration of the secure output space while maintaining functionality~\cite{fuConstrainedDecodingSecure2024}.

\sskip
Liang et al.~\cite{liangGrammarBasedCodeRepresentation2025} explore whether grammar-based code representations remain a worthy pursuit as language models scale to the billion-parameter level. While grammar-based methods were traditionally used to prevent syntax errors in smaller models, modern LLMs rarely make syntax mistakes. The authors developed \texttt{GrammarCoder}, a series of billion-scale models (1.a3B and~1.5B) that represent code as a sequence of grammar rules derived from the preorder traversal of an Abstract Syntax Tree (AST), rather than as a flat sequence of tokens. The paper's findings demonstrate that grammar-based models significantly outperform token-based baselines, even when both produce syntactically valid code. For instance, \texttt{GrammarCoder-1.3B-Base} showed an improvement of nearly seven percentage points in \emph{Pass@1} on the MBPP benchmark compared to \texttt{DeepSeek-Coder-1.3B-Base}. The authors attribute this gain to ``semantic differentiation.''

Their analysis reveals that grammar rules amplify representational differences for semantic shifts that appear minor at the token level; for example, neglecting operator precedence might change only two tokens but results in a~91\% increase in edit distance within a grammar-based representation. This sensitivity prevents the model from acting like a careless programmer, allowing it to better distinguish between logically distinct, but syntactically similar programs. The researchers conclude that explicitly capturing hierarchical structures and logical dependencies remains vital for improving the semantic accuracy and reliability of large-scale code models.

\sskip
Sistla et al.~\cite{sistlaVerifiedCodeReasoning2025} address the critical challenge of hallucinations in large language model agents tasked with reasoning about source code, which currently limits their utility in high-precision software engineering scenarios like code reviews or bug analysis. The authors propose a novel framework that automatically validates an agent's reasoning steps by extracting a \emph{formal representation} of its prose-based explanation, and verifying it using deterministic tools. This method introduces three key components: CodeSemantics to represent ground truth, AgentClaims to capture the formal version of the agent's response, and VerificationCondition to formally define the properties being checked. By converting natural language explanations into \emph{Datalog predicates}, using the Soufflé language, the system can leverage the efficiency and intuitive reasoning of LLMs, while also ensuring the final conclusion is mathematically sound and consistent with the underlying code.

The researchers evaluated their approach on two primary domains: uninitialized variable errors detected by sanitizers and program equivalence queries. For uninitialized variable tasks, the formal verifier successfully validated the agent's reasoning in thirteen out of twenty examples, though it required multiple iterations of predicate extraction to overcome initial inconsistencies in the model's output.

In the program equivalence domain, the framework effectively caught~75\% of the agent's hallucinations, identifying errors where the model made unwarranted assumptions about library behaviors or failed to consider critical control-flow dependencies. While the system still encounters challenges with incomplete predicate sets and complex code structures, the results demonstrate that post-facto formal verification significantly increases the trustworthiness of LLM-based code assistance by providing a rigorous layer of automated scrutiny.

\sskip
Tihanyi et al.~\cite{tihanyiNewEraSoftware2024} present a framework named ESBMC-AI that integrates Large Language Models (LLMs) with formal verification techniques to achieve automated software vulnerability repair. The primary motivation behind this work is the observation that while LLMs are proficient at generating alternative code solutions, they often lack the mathematical precision required to identify the root causes of vulnerabilities, particularly those involving bit-precise calculations or non-deterministic variables. Conversely, Bounded Model Checking (BMC) provides rigorous mathematical proofs of property violations but requires significant expertise to interpret and manually fix the resulting counterexamples. ESBMC-AI bridges this gap by using the Efficient SMT-based Context-Bounded Model Checker  to identify vulnerabilities and generate detailed stack traces and counterexamples.

These counterexamples, which specify line numbers and error types, are combined with the original source code into a specialized prompt for an LLM. The model is then instructed to repair the code based on this formal guidance. The resulting patch is automatically re-verified by ESBMC to ensure that the fix is not only syntactically correct but also formally safe. This iterative process continues until the verification is successful. The researchers evaluated the framework on over~1,000 C programs from the FormAI dataset, achieving impressive repair accuracies:~90.40\% for buffer overflows on \emph{scanf},~86.47\% for division by zero, and approximately~70\% for arithmetic overflows and array bounds violations. The study also addresses the undecidability of formal verification by using cyclomatic complexity measurements to ensure that the generated patches do not drastically alter the program's intended workflow. This integration of symbolic execution and generative AI offers a scalable pathway for integrating ``self-healing'' capabilities into modern CI/CD pipelines.

\sskip
Meijer~\cite{meijerGuardiansAgents2026} introduces a safety paradigm for agentic applications, which are AI systems empowered to take autonomous actions by calling external tools. The author argues that giving autonomous agents access to tools with potentially irreversible side effects carries significant risks, such as reward hacking or falling victim to prompt injection attacks. These risks are exacerbated by the current inability of models to reliably distinguish between instructions and data, allowing malicious actors to coerce agents into harmful actions. To mitigate these risks, Meijer proposes a shift from reactive monitoring to a ``secure-by-construction'' approach rooted in mathematical proof verification.

In this design pattern, an AI agent must generate a formal proof demonstrating the safety of its planned actions before they are authorized for execution. This approach is compared to real-world mechanisms like banknote security features, where the complexity lies in the production of the artifact, but verification is simple and efficient. By utilizing static verification, the system can ensure that policy violations are impossible before any action is taken, eliminating the need for complex rollback mechanisms. Meijer emphasizes that this method does not require trusting the AI itself; instead, it relies on a deterministic and verifiable process to provide robust assurances. The paper also discusses the limitations of traditional guardrails, such as false positives and cultural biases, and contrasts them with the proposed verification-based model, which offers a more scalable and automated way to integrate security into the workflow-generation pipeline.

\sskip
Shi and Zhang~\cite{shiRESCUERETRIEVALAUGMENTED2025} propose \textsc{Rescue}, a framework that enhances Retrieval-Augmented Generation (RAG) for secure code generation. They identify that conventional RAG often fails in the security domain because raw security documents contain significant noise and irrelevant logic that distracts the LLM. To overcome this, \textsc{Rescue}  introduces a hybrid knowledge base construction method. First, it uses an LLM-assisted ``cluster-then-summarize'' pipeline to distill high-level, actionable security guidelines from raw vulnerability-fix data. Second, it applies static program slicing to extract concise, security-focused code examples, removing all logic unrelated to the security fix. The framework then utilizes a hierarchical multi-faceted retrieval strategy. This strategy performs a coarse search for relevant CWE types followed by a fine-grained search across three facets: API patterns, vulnerability cause analysis, and code structure.

By fusing these multi-faceted results, \textsc{Rescue}  provides the LLM with a highly relevant security context tailored to the specific task description. In evaluations across four benchmarks and six LLMs, \textsc{Rescue}  improved security metrics (SecurePass@1) by an average of~4.8 points while preserving functional correctness. The authors demonstrate that distilling and structuring security knowledge is essential for making RAG an effective tool for preventing the generation of vulnerable code.

\sskip
Li et al.~\cite{liVULSOLVERVulnerabilityDetection2025} present \textsc{VulSolver}, a framework that reframes vulnerability detection as a path-based constraint-solving problem. The authors argue that traditional LLM-based detection is often unstable and prone to hallucinations because it tries to map code directly to a label without structured reasoning. \textsc{VulSolver}  instead mimics human security experts by decomposing the analysis of a potential vulnerability path into ``transfer constraints'' and ``trigger constraints.'' Transfer constraints verify whether a sink point is reachable from a source via a feasible execution path, while trigger constraints determine if the input reaching that sink can actually satisfy the conditions for exploitation (e.g., whether a path parameter can contain \texttt{..} for a path traversal attack).

The system uses static analysis to extract candidate call paths and then prompts the LLM to solve these constraints progressively for each caller-callee pair. This structured approach allows the model to process code in a controlled manner, focusing on localized logic while building upon prior conclusions. Evaluated on the OWASP Benchmark, \textsc{VulSolver}  achieved~100\% recall and~97.85\% accuracy. More importantly, it discovered 15 previously unknown high-severity vulnerabilities in popular open-source projects. The authors highlight that by acting as a semantic constraint solver, the LLM provides more robust and verifiable results than simple classification models.

\subsection{Alternative Representations for Code Generation}
As argued in recent position papers, moving from probabilistic pattern matching to truly reliable software artifacts necessitates the use of \emph{structured representations} similar to compiler intermediate representations~(IRs), formal correctness frameworks, and robust verification mechanisms. Future research in this domain likely involves designing new LLM architectures or training objectives that explicitly teach code generation models to do formal reasoning and have them at least partially simulate execution behavior.

\point{Syntactic and Efficient Constrained Decoding}
A primary challenge in ensuring LLMs generate structured output is token misalignment, where subword tokens used by model vocabularies do not align directly with the terminals of a formal grammar. Early attempts at constrained decoding often incurred high performance overhead or impaired task accuracy by failing to correctly align these vocabularies. \textsc{Domino}~\cite{beurer-kellnerGuidingLLMsRight2024} addresses this by providing a minimally invasive decoding algorithm that enforces constraints in a fully subword-aligned fashion. It leverages pre-computation and a novel speculative decoding procedure to achieve virtually no overhead, and in some cases, significant speedups over unconstrained decoding. \textsc{Domino} defines a minimally invasive method as one where every valid output that can be generated by an unconstrained model is also generated by the constrained model for the same prompt.
While \textsc{Domino} focuses on inference \emph{speed}, \textsc{GreatGamma}~\cite{parkFlexibleEfficientGrammarConstrained2025} addresses the prohibitive offline preprocessing costs associated with many grammar-constrained decoding~(GCD) algorithms, which can take tens of minutes for common grammars. \textsc{GreatGamma} introduces a token spanner table that efficiently precomputes a lexer-state-dependent mapping between sequences of context-free grammar~(CFG) tokens and individual LLM tokens. This approach achieves an average~$17.71\times$ speedup in offline preprocessing, while maintaining state-of-the-art online masking efficiency, making it practical for domains where the constraining grammar changes frequently, such as program synthesis or grammar prompting.
We can envision techniques that use more complex IR and grammars for creating secure-by-construction code output.

\point{Semantic Constrained Decoding and Formal Verification}
Syntactic constraints alone are often insufficient for mission-critical applications, as they cannot enforce properties like type safety or functional equivalence. \textsc{ChopChop}~\cite{nagyChopChopProgrammableFramework2025} introduces the first programmable framework for semantically constraining the output of language models. It formulates the analysis of program spaces as a realizability problem solved via coinduction, connecting token-level generation with structural reasoning over programs. By representing program spaces as regular codata and using corecursive semantic pruners, \textsc{ChopChop} can enforce complex properties such as equivalence to a reference program and type safety during the decoding process.

\point{Neuro-Symbolic Execution and Modeling Computation}
Beyond external constraints, researchers have sought to embed computational logic directly into neural models. Neural Interpretation~(NI) is a neuro-symbolic execution model that reads Python source code and executes it statement-by-statement using neural networks composed according to the code structure~\cite{huNeuroSymbolicExecutionGeneric2023}. NI formulates the Neuro-Symbolic Execution Problem (NSXP) as learning an isomorphic mapping between source code and neural representation spaces. Every variable is modeled as a latent vector (abstract semantics), and every function executes a neural network (the Executor). Unlike standard LLMs that process code as linear token sequences, NI preserves source code compositionality and can perform white-box execution even with missing definitions or library dependencies. This allows it to localize and repair variable misuses more effectively than purely statistical baselines.

\point{Uncertainty Estimation and the Semantic Gap}
A critical aspect of reliable code generation is the model's ability to assess its own correctness. In natural language, information-theoretic uncertainty techniques (e.g., semantic entropy) are used as proxies for quality, but these often fail when applied to code because existing embeddings and heuristic metrics like \textsc{CodeBLEU} cannot reliably capture functional equivalence. Sharma and David~(2025)~\cite{sharmaAssessingCorrectnessLLMBased2025} propose symbolic clustering to address this, using symbolic execution to group semantically equivalent programs even when they vary significantly in syntax. By integrating symbolic clustering into uncertainty frameworks, they restore the predictive power of these techniques, showing that semantic equivalence is a more dominant signal for correctness than probabilistic confidence. They also introduce the symbolic cluster count as a lightweight correctness proxy that supports highly effective abstention policies.

\subsection{Diffusion Models}
\label{sec:diffusion-models}
Zeng et al.~\cite{zengTreeDiffASTGuidedCode2025} addresses the limitations of applying standard diffusion-based language models to the highly structured domain of source code. While traditional diffusion models use random token-level masking, the authors argue that this ignores the strict syntactic and hierarchical rules inherent in programming languages. To resolve this, they introduce TreeDiff, a syntax-aware diffusion framework that incorporates structural priors from Abstract Syntax Trees~(ASTs) directly into the corruption process. Instead of random masking, TreeDiff selectively masks tokens belonging to key AST nodes or subtrees. This strategy encourages the model to internalize the compositional nature of code, enabling it to reconstruct programs that respect grammatical boundaries and capture long-range dependencies.
Evaluated across multiple benchmarks, including HumanEval and MBPP, TreeDiff consistently outperformed standard random masking strategies. Notably, it achieved a~13.3\% relative improvement on HumanEval+, demonstrating the effectiveness of leveraging underlying structural scaffolds for modeling code. The work highlights that syntax-guided denoising is a promising alternative to autoregressive models, offering better reconstruction accuracy and generalization to unseen code patterns.

Kapur et al.~\cite{kapurDiffusionSyntaxTrees2024} propose a novel approach for program synthesis by moving the diffusion process entirely onto the syntax trees of context-free grammars (CFGs). Recognizing that autoregressive models often lack the feedback required to observe a program's output during generation, the authors develop neural diffusion models that iteratively edit code while preserving syntactic validity. The ``forward'' process involves making small, random, yet syntactically valid mutations to a syntax tree, while the "reverse" process trains a conditional neural network to predict the edit path back to the target program.
A key technical contribution is the use of a tree edit path algorithm based on tree edit distance, which provides a cleaner training signal than simply inverting individual mutations. The authors apply this framework to inverse graphics tasks, where the model learns to convert images or hand-drawn sketches into the code required to reproduce them. By combining the policy network with a value network to guide beam search, the system can "see" the execution results and debug code to meet required specifications. This work demonstrates that iterative editing on trees is a viable and efficient alternative to sequential token generation.

\newcommand{\DiffuGuard}{\textsc{DiffuGuard}\xspace}

As Diffusion Large Language Models~(dLLMs) advance, Li et al.~\cite{liDiffuGuardHowIntrinsic2025} introduce unique security vulnerabilities fundamentally different from those in autoregressive models. In DiffuGuard, Li et al. conduct the first systematic analysis of dLLM vulnerabilities to jailbreak attacks, focusing on ``intra-step'' and ``inter-step'' dynamics. Their research identifies a critical phenomenon called \emph{Denoising-path Dependence}, where the safety of early-stage tokens decisively influences the final output. They discover that the standard greedy remasking strategy used in dLLMs creates a harmful bias that amplifies the selection of malicious content.
To mitigate these risks, the authors propose \DiffuGuard, a training-free defense framework. It utilizes \emph{Stochastic Annealing Remasking} to introduce controlled randomness into the decoding process, breaking the harmful paths created by greedy selection. Additionally, a block-level audit and repair mechanism uses internal model representations to detect and correct unsafe segments during generation. Experiments across four dLLMs showed that \DiffuGuard reduced the Attack Success Rate (ASR) from~47.9\% to~14.7\% without significantly impacting general model utility or efficiency.


\newcommand{\CodeDiffuSe}{\textsc{CodeDiffuSe}\xspace}

Onan and Alhumyani~\cite{onanCodeDiffuSeMaskedDiffusion2025} introduce \CodeDiffuSe, a masked diffusion framework specifically designed for code completion and bug repair. Traditional autoregressive models often struggle with ``infilling'' tasks where missing or buggy code spans are located in the middle of a sequence. \CodeDiffuSe overcomes this by leveraging global bidirectional context, allowing it to reason about both left and right surrounding code simultaneously. The framework employs a syntax-aware masking strategy that targets entire AST subtrees during training, alongside a semantic consistency regularization that encourages the generation of type-correct and syntactically valid code.
During inference, \CodeDiffuSe uses an error-aware remasking mechanism to dynamically identify and refine uncertain or invalid tokens across adaptive reverse diffusion steps. Extensive evaluation on benchmarks such as CodeXGLUE, Defects4J, and QuixBugs showed that \CodeDiffuSe consistently outperformed strong autoregressive baselines like CodeGen and InCoder. The results particularly highlight gains in AST accuracy and compilation success rates, establishing the framework as a robust, structure-aware alternative for complex software engineering tasks.

Li et al.~\cite{liAutoregressionEmpiricalStudy2025} present the first comprehensive empirical study of Diffusion LLMs for code generation, evaluating nine representative dLLMs across four major benchmarks. The authors highlight two primary advantages of dLLMs: multi-token prediction (parallel decoding), which significantly improves inference speed, and flexible generation order, which allows for iterative refinement of any part of the sequence.
The paper's findings reveal that existing dLLMs are highly competitive with AR models of similar size. For example, on HumanEval and MBPP, the best dLLMs achieved pass@1 scores that outperformed their AR counterparts. Additionally, the authors found that dLLMs possess superior length extrapolation abilities and perform better at understanding long-range dependencies in complex codebases. By providing practical guidance on the factors impacting dLLM effectiveness and efficiency, this work challenges the assumption that autoregressive modeling is the only viable path for scalable code intelligence.

Zhang et al.~\cite{zhangExploringPowerDiffusion2025} present a large-scale evaluation of Diffusion LLMs (DLLMs) across the entire software development lifecycle~(SDLC), including code generation, defect detection, program repair, and cross-file maintenance. They argue that traditional AR-LLMs are limited by high inference latency and a causal masking mechanism that prevents them from simultaneously observing long-range hierarchical relationships. In contrast, DLLMs utilize global bidirectional encoding and ``step-length decoupling,'' where the number of refinement iterations is independent of the sequence length.
Evaluating~52,937 tasks, the researchers found that~7B-parameter DLLMs outperformed AR-LLMs of similar scale with an average accuracy improvement of~30\%. The most significant gain was observed in cross-file repair, where DLLMs achieved a~113\% improvement over AR models. These results suggest that the ``global refinement'' paradigm of diffusion models is naturally better suited for real-world software engineering tasks that require deep context awareness and low-latency interaction. This investigation establishes DLLMs as a superior and more efficient paradigm for automated program generation and daily maintenance.

\subsection{Guardrails for Security}
\sskip
Roychoudhury et al.~\cite{roychoudhuryAgenticAISoftware2025} explore the transition from ``programming in the large'' to ``programming with trust'' in the context of AI software engineers. While LLMs have demonstrated proficiency in generating code snippets, the authors argue that their deployment in industrial practice is hindered by a lack of trust among developers. To bridge this gap, they envision the use of LLM agents—wrappers around LLMs that interact with software testing and analysis tools to enhance the reliability and maintainability of generated code. These agents are characterized by their ability to autonomously create work plans and utilize tools like file navigators, test suites, and program analysis engines.

The authors highlight several mechanisms for establishing trust, including the generation of tests alongside code, the inference of software intent from natural language specifications, and the use of formal proofs. Specifically, the paper advocates for ``intent inference,'' where agents interpret program representations (such as Abstract Syntax Trees) to make micro-decisions that are more explainable and accurate than surface-level text processing. Furthermore, the researchers propose the implementation of guardrails as sanitization mechanisms to filter malicious inputs, such as prompt injections, and to validate outputs for security vulnerabilities. By integrating these technical and human-centric trust-building measures, the paper suggests that AI software engineers can eventually co-exist with human developers in future workflows. The authors conclude that as programming becomes less about writing and more about assembling components, the focus of software engineering must shift toward ensuring the correctness and security of these assemblies through agentic oversight.

\sskip
Christodorescu et al.~\cite{christodorescuSystemsSecurityFoundations2025} articulate the foundational research problems in AI agent security through the lens of computer systems security. The authors argue that hardening individual AI models (AI alignment) is insufficient because it mirrors the failed historical approach of trying to write ``perfect'' software. Instead, they advocate for a systems-level approach that examines end-to-end security properties, incorporating deterministic guardrails at various layers of abstraction. This approach acknowledges that the AI model is a probabilistic component and should be treated as part of a larger, untrusted system. The paper defines the essential elements of a security architecture for agentic computing: the Trusted Computing Base (TCB), security policies, and security boundaries.

A major challenge identified is the ``Probabilistic TCB,'' where the security monitor—often another model—is itself fundamentally uncertain and conditional on input, making it difficult to enforce invariants with~100\% reliability. Additionally, the authors discuss the ``Fuzzy Security Boundary,'' noting that agents often operate at low levels (like UI clicks) where it is difficult to define and enforce meaningful semantic policies. The ``Dynamic Security Policy'' problem is also explored, highlighting that because agents ``compute'' their own programs based on underspecified natural language tasks, there is often no clear developer-defined intent to analyze. To illustrate these concepts, the paper presents 11 case studies of real-world attacks on agentic systems, demonstrating how traditional principles like the \emph{Principle of Least Privilege} and \emph{Complete Mediation} can be adapted to stop modern exploits. The work provides a rigorous framework for practitioners to move beyond ad-hoc safety patches toward a robust, systems-oriented security posture for autonomous AI agents.

\sskip
Dai et al.~\cite{daiRethinkingEvaluationSecure2025} challenge the prevailing evaluation methodologies for secure code generation, arguing that current metrics often overlook the trade-off between security and functional correctness. Most existing studies evaluate these two aspects independently, using security-specific datasets for vulnerability checks and general datasets like HumanEval for functionality. This separation creates a blind spot: a model might pass security tests by generating ``garbage code'' that is technically safe because it does nothing, or by entirely removing vulnerable lines of code that were necessary for the program's logic. To address this, the authors introduce the \emph{SAFE} metric, which evaluates both security and functionality on the same piece of generated code, providing a more holistic view of model performance.

The researchers applied this combined measure to four state-of-the-art secure code generation techniques---\textsc{SVEN}, \textsc{SafeCoder}, \textsc{CodeGuard+}, and \textsc{PromSec}---using large-scale datasets like BigCodeBench and \textsc{SecCodePLT+}. Their analysis reveals a disturbing trend: many techniques significantly degrade the base model's performance, sometimes by more than~50\%, while only offering marginal security improvements. Furthermore, the study identifies a ``security overestimation'' problem caused by relying on a single static analyzer like \textsc{CodeQL}. By employing three different analyzers and two LLMs for detection, the authors found that \textsc{CodeQL} alone misses more than~20\% of vulnerabilities in generated code. The paper serves as a critical call to action for the research community to adopt more rigorous, multi-faceted evaluation schemes that prevent models from \emph{cheating} the security metric at the expense of utility. They emphasize that a truly secure code generator must maintain the integrity of the original task while enforcing safety constraints.

\sskip
Wei et al.~\cite{weiSmartAuditFlowDynamicPlanExecute} propose SmartAuditFlow, a dynamic ``Plan-Execute'' framework designed to enhance the reliability and accuracy of smart contract security auditing using Large Language Models. Smart contracts represent a unique security challenge due to their immutable nature and the high financial stakes of decentralized finance. While LLMs show promise in code analysis, they suffer from hallucinations and a limited ability to perform complex, multi-step reasoning. SmartAuditFlow addresses these issues by decomposing the auditing task into five sequential stages: context-aware initial analysis, adaptive audit planning, multi-faceted vulnerability execution, cross-cutting findings synthesis, and comprehensive report generation.

The framework features a Prompt Optimization Engine that iteratively refines the instructions given to the LLM based on real-time feedback, ensuring that the model remains focused on the contract's specific business logic and risk profile. Additionally, SmartAuditFlow integrates Retrieval-Augmented Generation (RAG) and external static analysis tools to ground the model's reasoning in up-to-date security patterns and authoritative knowledge bases. In experimental evaluations using metrics like Mean Reciprocal Rank and Mean Average Precision, the framework demonstrated significant improvements over traditional static tools and baseline LLM implementations. It successfully identified~13 additional CVEs that were missed by existing methods, showing high precision in detecting subtle logic errors and inter-contract vulnerabilities. The authors conclude that by emulating the methodical, iterative approach of human expert auditors, SmartAuditFlow provides a scalable and robust solution for ensuring the security of blockchain-based applications before deployment.

\sskip
Zhang et al.~\cite{zhangLowCostComprehensiveNontextual2025} introduce, a novel framework that leverages Large Language Models to augment greybox fuzzing for software that accepts non-textual inputs, such as images, videos, and PDF files. Traditional fuzzers often struggle with these formats because they require a deep understanding of complex grammars that are often unavailable or hard to infer. While LLMs excel at generating textual data, they are often incapable or prohibitively expensive when asked to generate binary files directly.  overcomes this by using LLMs to synthesize and mutate \emph{input generators}---Python scripts that utilize existing libraries to produce data conforming to the target grammar.

The framework employs a hybrid strategy that combines a \emph{holistic search} driven by LLMs with a ``local search'' driven by industrial-quality fuzzers like AFL++. The LLM is used to synthesize diverse generators that explore high-level structural variations, effectively helping the fuzzer ``jump out'' of local optima. These generators then produce seed inputs that are rapidly mutated at the byte level by AFL++ to find deeper bugs. This synergistic approach significantly reduces the cost of LLM usage; in one~24-hour campaign, the cost of GPT-3.5 API calls was less than \$0.20. The researchers evaluated  on 34 different input formats across platforms like FuzzBench and MAGMA, consistently outperforming state-of-the-art tools in terms of code coverage and bug finding. The framework discovered 10 unique bugs in real-world software, including three new CVEs, demonstrating that using LLMs to generate the logic for test case creation is a highly effective way to handle the complexity of modern binary formats.

\sskip
Zhu et al.~\cite{zhuSpecificationGuidedVulnerabilityDetection2025} present \textsc{VulInstruct}, a specification-guided approach for vulnerability detection that addresses the inability of standard LLMs to distinguish between vulnerable and patched code. The authors argue that LLMs primarily rely on surface-level pattern matching because they lack an understanding of ``security specifications''~---~the implicit expectations defined by developers about how code should behave to remain safe. For example, a TLS implementation might appear functional but is insecure if it lacks hostname verification. VulInstruct systematically extracts these reusable specifications from historical vulnerabilities and high-quality patches to create a knowledge base.

The framework utilizes two automatic pipelines to construct this base: one for general specifications derived from the open-source ecosystem and another for domain-specific expectations repeatedly violated in particular codebases. When analyzing new code, \textsc{VulInstruct} retrieves relevant past cases and their associated specifications, providing the LLM with the explicit security knowledge needed to reason about root causes. This approach was evaluated using the \textsc{PrimeVul} dataset and the strict CORRECT evaluation framework, which requires models to provide both a correct label and a valid reasoning trace. \textsc{VulInstruct} achieved a~45.0\% F1-score, a~32.7\% improvement over the strongest baselines, and was particularly effective at detecting unique vulnerabilities that other methods missed entirely. The practical value of the tool was further demonstrated by its discovery of a previously unknown high-severity vulnerability in production code (CVE-2025-56538). This work highlights that providing models with the `why'' behind security rules is essential for moving beyond simple binary classification to true semantic understanding~\cite{zhuSpecificationGuidedVulnerabilityDetection2025}.

\sskip
Kharma et al.~\cite{kharmaSecurityQualityLLMGenerated2025} conduct a comprehensive multi-language and multi-model analysis of the security and maintainability of code generated by state-of-the-art LLMs, including Claude-3.5, Gemini-1.5, and GPT-4o. The researchers introduced a manually vetted dataset of~200 programming tasks across four languages—Python, Java, C++, and C—to investigate how language selection and model architecture influence the quality of generated code. A key finding of the study is that LLM security effectiveness varies significantly by programming language; for instance, many models fail to utilize modern security features available in recent updates like Java~17, opting instead for outdated or insecure practices.

The study utilized static security analysis tools to identify common vulnerabilities, such as improper buffer handling in C++ and lack of defensive programming in Python. The authors observed that LLMs often display overconfidence in the security of their outputs, generating code that is functionally correct but lacks robust error handling or sanitization. Furthermore, the analysis reveals that models often inherit the \emph{vulnerability profile} of their training data, frequently repeating common historical mistakes. The paper evaluates several quality metrics, including validity, consistency, and responsibility, noting that while model size generally correlates with better quality, it does not always guarantee safer code. The researchers emphasize the need for advancing LLMs to incorporate emerging best practices and \emph{security personas} to guide generation. This work provides a detailed benchmark for comparing the security posture of modern AI assistants and highlights the critical need for language-specific security guidelines in the era of automated software development.

\cleardoublepage
\bibliographystyle{plain}
\bibliography{biblio}

\end{document}